\documentclass[12pt,letterpaper]{article}
\usepackage[a4paper, total={7in, 10in}]{geometry}

\usepackage{graphicx}
\usepackage{helvet}
\usepackage{authblk}
\usepackage{hyperref}
\usepackage{amsmath} 
\usepackage{amssymb} 
\usepackage{orcidlink} 
\usepackage[super,comma,sort&compress]  
   {natbib}
\usepackage[right]{lineno} \linenumbers
\usepackage[title]{appendix}%

\usepackage[utf8]{inputenc} 
\usepackage[T1]{fontenc}    
\usepackage{url}            
\usepackage{booktabs}       
\usepackage{amsfonts}       
\usepackage{nicefrac}       
\usepackage{microtype}      
\usepackage{xcolor}         
\usepackage{multirow}
\usepackage{threeparttable}
\usepackage{amsmath}
\usepackage{subcaption}
\usepackage{makecell}
\usepackage{placeins}
\usepackage[utf8]{inputenc} 
\usepackage{siunitx}       
\usepackage{caption}       
\usepackage{tabularx}
\usepackage{listings}    
\usepackage{pifont}
\usepackage{makecell}
\usepackage{longtable}
\usepackage{booktabs}
\usepackage{colortbl}
\usepackage[table]{xcolor}
\usepackage{tikz}
\usepackage{array}
\usepackage{geometry}
 
\definecolor{groupgray}{RGB}{243,243,241}
\definecolor{totalblue}{RGB}{232,241,252}
\definecolor{ciblue}{RGB}{25,95,190}
\definecolor{citeal}{RGB}{12,115,95}

\definecolor{sub}{gray}{0.42}
 
\newcolumntype{L}[1]{>{\raggedright\arraybackslash}p{#1}}
\newcolumntype{R}[1]{>{\raggedleft\arraybackslash}p{#1}}

\theadset{\bfseries}

\makeatletter
\renewcommand{\maketitle}{\bgroup\setlength{\parindent}{0pt}
\begin{flushleft}
  \textbf{\@title}
  
  \@author
\end{flushleft}\egroup}
\makeatother

\title{Wearable Single-Lead ECG Detects Fine-Grained Structural Heart Disease Through Echo-Report Supervision}
\date{}

\author[1,$\#$]{Chenyang He} 
\author[2,$\#$]{Qinghao Zhao} 
\author[1]{Shun Huang} 
\author[1,3]{Jun Li} 
\author[1,3]{Gongzheng Tang} 
\author[4]{Hao Zhang} 
\author[4]{Tong Liu} 
\author[4]{Zhengkai Xue} 
\author[2]{Jian Liu} 
\author[4,*]{Kangyin Chen} 
\author[5,*]{Cheng Ding} 
\author[1,3,6,7,*,\orcidlink{0000-0001-7521-5127}]{Shenda Hong} 

\affil[1]{National Institute of Health Data Science, Peking University, Beijing, China}
\affil[2]{Department of Cardiology, Peking University People’s Hospital, Beijing, China}
\affil[3]{Institute of Medical Technology, Peking University Health Science Center, Beijing, China}
\affil[4]{Tianjin Key Laboratory of lonic-Molecular Function of Cardiovascular Disease, Department of Cardiology, Tianjin Institute of Cardiology, The Second Hospital of Tianjin Medical University,Tianjin, China}
\affil[5]{Nanjing University of Aeronautics and Astronautics, Nanjing, China}
\affil[6]{Institute for Artificial Intelligence, Peking University, Beijing, China}
\affil[7]{State Key Laboratory of Vascular Homeostasis and Remodeling, NHC Key Laboratory of Cardiovascular Molecular Biology and Regulatory Peptides, Peking University, Beijing, China}

\affil[$\#$]{These authors contributed equally}
\affil[*]{Correspondence: hongshenda@pku.edu.cn, chengding@nuaa.edu.cn, chenkangyin@vip.126.com}

\begin{document}

\maketitle

\section*{ABSTRACT}

Structural heart disease (SHD) is a primary driver of heart failure and cardiovascular mortality, yet early detection remains constrained by the limited accessibility of echocardiography. While single-lead electrocardiogram (ECG) is ubiquitous through wearables, existing AI screening models often depend on 12-lead inputs, generalize poorly across institutions, or require massive, condition-specific labeled datasets. Recent work has demonstrated the feasibility of contrastive pre-training between single-lead ECGs and echocardiography reports within a single health system. Here, we present AnyECG-Echo, a framework that advance this paradigm toward clinical translation through three key developments: (1) evaluation in a geographically independent external cohort ($n = 16,621$); (2) diagnostic coverage of 13 fine-grained SHD subtypes spanning myocardial, chamber, valvular, and great-vessel pathologies; and (3) dual-axis mechanistic interpretability combining electrophysiology-grounded Shapley attribution with emergent correlations to quantitative measurements. Across validation cohorts totaling $n = 25,222$, the model demonstrated high AUROC for high-impact subtypes, including reduced left ventricular systolic function (AUROC 0.866–0.924), global heart enlargement (0.877–0.931), and mitral stenosis (0.836–0.906). Furthermore, we successfully validated the alignment of model outputs with established medical physiological traits, thereby enhancing interpretability. Notably, we discovered that AnyECG-Echo’s outputs function as physiologically grounded digital biomarkers that accurately track objective metrics such as $LVEF$ and myocardial wall thickness. These findings prove that wearable single-lead ECGs can effectively detect fine-grained structural heart disease, offering a practical solution for population-scale screening.

\section*{KEYWORDS}
Structural Heart Disease, Multimodal Fusion, Single-lead Electrocardiogram

\section*{INTRODUCTION}

Structural heart disease (SHD)~\cite{dhingra2025ensemble,aminorroaya2025development}, encompassing myocardial~\cite{taggart2025targeting}, chamber~\cite{ortiz2025double}, valvular~\cite{unger2025mixed}, and great-vessel pathologies~\cite{yao2023graph}, is a primary driver of cardiovascular mortality worldwide. A defining challenge of SHD is its prolonged, clinically silent prodrome~\cite{lee2023structural}. By the time overt symptoms manifest, irreversible cardiac remodeling has often occurred, rendering disease-modifying therapies less effective. Consequently, improving population-level outcomes hinges on early detection through scalable screening infrastructure~\cite{antoniades2025scalable}. While echocardiography remains the diagnostic reference standard~\cite{hickson2016echocardiography}, its reliance on specialized equipment and certified operators creates a profound clinical bottleneck~\cite{jiang2025ultrasep}. These constraints concentrate diagnostic capacity within tertiary centers, leaving much of the early SHD burden invisible to health systems until structural compromise is advanced~\cite{krammel2023feasibility}.

The electrocardiogram (ECG) represents the only cardiac modality available at population scale. It is inexpensive, non-invasive, and increasingly accessible through consumer wearables~\cite{palermi2025artificial,alimbayeva2022portable,wang2024systematic,lin2026self}. However, three major barriers impede its clinical translation for SHD screening~\cite{zhu2026artificial}. First, most high-performance models are optimized for 12-lead configurations, incompatible with the single-lead sensors used in consumer wearables~\cite{poterucha2025detecting,hughes2026echonext}. Second, variations in hardware and demographics across healthcare sites frequently compromise model generalizability. Third, the prevailing supervised-learning paradigm relies on massive, manually labeled datasets for each specific condition, which are expensive to assemble and lack diagnostic granularity~\cite{liang2025artificial}.

Recent work has begun exploring contrastive pre-training between single-lead ECGs and echocardiography reports as a means to bypass manual labeling entirely~\cite{Wearable-Echo-FM}. While these efforts have established the feasibility of this paradigm within single health systems, three critical gaps remain before clinical deployment: the absence of evidence for cross-institutional generalization across different geographies and clinical practices; limited diagnostic granularity, with current frameworks targeting only a small set of coarse subtypes (e.g., left ventricular systolic dysfunction, composite SHD); and the absence of mechanistic verification that learned representations reflect cardiac physiology rather than center-specific spurious correlations. To address these challenges, we developed AnyECG-Echo, a multimodal contrastive learning framework that repurposes routinely generated, physician-authored echocardiography reports as an automated and high-fidelity supervisory signal. We systematically validated this framework across three distinct validation tiers: an internal testing set ($n = 4,078$), a temporal test set ($n = 4,523$), and a geographically independent external test set ($n = 16,621$).

Crucially, we interrogated the model’s clinical validity through two complementary lenses: first, by aligning ECG-segment attribution with expert-defined electrophysiological signatures, and second, by demonstrating that the model's representations exhibit correlations with raw quantitative measurements never encountered during training. These findings suggest that the model's output functions as a digital biomarker of SHD. 
Collectively, these analyses demonstrate that AnyECG-Echo provides a scalable, mechanistically grounded foundation for SHD pre-screening. By bridging the gap between ubiquitous sensor technology and definitive diagnostics, this framework enables expert-level assessment in out-of-hospital environments, remote communities, and primary care settings where echocardiography remains inaccessible.

\begin{figure*}[!]
    \centering
    \includegraphics[width=1.0\textwidth]{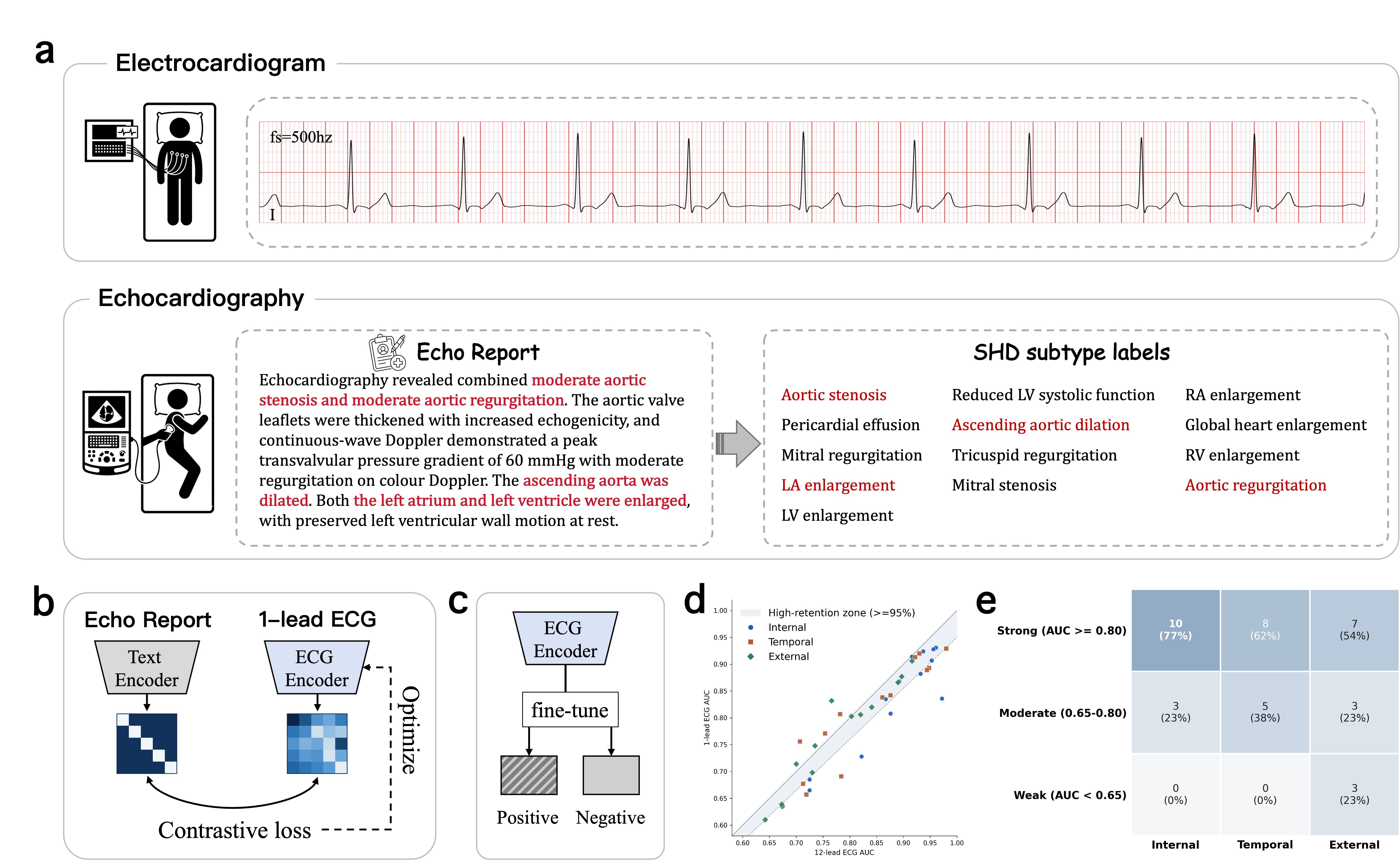}
    \caption{\textbf{Overview of the AnyECG-Echo framework and its performance across heterogeneous evaluation cohorts.}
    \textbf{(a)} Schematic of the multimodal input pipeline. A single-lead ECG recording (Lead I, sampled at 500 Hz) is paired with its corresponding physician-authored echocardiography report, from which SHD subtype labels are extracted.
    \textbf{(b)} Multimodal contrastive pre-training stage. An ECG encoder and a text encoder process paired ECG-echocardiography report inputs. Their representations are aligned via a contrastive loss to facilitate the transfer of structural cardiac semantics into the ECG embedding space.
    \textbf{(c)} Supervised fine-tuning stage. With the pre-trained ECG encoder parameters frozen, a lightweight linear classification head is optimized to distinguish between positive and negative cases for each SHD label.
    \textbf{(d)}, Performance retention from 12-lead to 1-lead ECG across structural heart disease labels.
    Each point represents one label in one cohort (internal, temporal and external), with the x-axis denoting 12-lead AUROC and the y-axis denoting 1-lead AUROC. The shaded band indicates high retention ($\geq$95\%).  
    \textbf{(e)}, Distribution of 1-lead predictive performance tiers across cohorts. Rows indicate predefined AUROC tiers (Strong, Moderate and Weak), and columns indicate cohorts.}
    \label{main}
\end{figure*}

\section*{RESULTS}

\subsection*{AnyECG-Echo achieves high-fidelity SHD screening with single-lead inputs}

AnyECG-Echo bridges the gap between accessible electrophysiology and definitive structural imaging by aligning single-lead ECG (Lead I) embeddings with the semantic narratives of physician-authored echocardiography reports (Fig.~\ref{main}a–c). By leveraging these reports as a dense, automated supervisory signal, the framework internalizes the structural implications of electrical signals across 13 fine-grained subtypes, effectively eliminating the requirement for manual expert annotation.

A critical concern in wearable-based screening is the potential diagnostic loss incurred when reducing 12-lead configurations to a single lead. We quantified this trade-off by calculating the AUROC retention ratio (single-lead AUROC divided by 12-lead AUROC). As shown in Fig.~\ref{main}d, the majority of conditions exhibited high retention ($\geq 95\%$), with 93\% of all condition–cohort combinations exceeding the 90\% threshold. Notably, for high-impact categories such as reduced LV systolic function and right atrial enlargement, single-lead performance was nearly identical to the 12-lead baseline (retention $> 97\%$). These findings demonstrate that Lead I captures the primary electrophysiological signatures of SHD with sufficient fidelity to support clinical-grade screening.

Across all three validation tiers, AnyECG-Echo demonstrated reliable discriminative capacity for the 13 fine-grained SHD subtypes (Fig.~\ref{main}e). The majority of targeted subtypes achieved strong discrimination ($\text{AUROC} \ge 0.80$). While Aortic regurgitation, Ascending aortic dilation, and LA enlargement fell within a weaker range ($\text{AUROC} < 0.65$) in the external test set, the vast majority of categories across all cohorts maintained acceptable to strong predictive power ($\text{AUROC} \ge 0.65$), confirming the framework's screening utility.

\begin{table}[htbp]
  \centering
  \begin{threeparttable}
    \caption{Model Performance Evaluation Across Different Test Sets}
    \label{tab:combined_performance}
    \footnotesize
    \begin{tabular}{lcccc}
      \toprule
      Field ID & AUROC (95\% CI) & Sensitivity & Specificity & Brier Score \\
      \midrule
      \multicolumn{5}{l}{\textbf{Internal Test Set from PKUPH}} \\
      \quad Global heart enlargement & 0.931 $[0.880, 0.971]$ & 0.810 & 0.933 & 0.005 \\
      \quad Aortic regurgitation     & 0.928 $[0.841, 0.979]$ & 0.850 & 0.899 & 0.004 \\
      \quad Reduced LV systolic fn.  & 0.924 $[0.895, 0.950]$ & 0.821 & 0.905 & 0.021 \\
      \quad RA enlargement           & 0.914 $[0.871, 0.952]$ & 0.868 & 0.837 & 0.008 \\
      \quad Tricuspid regurgitation  & 0.907 $[0.838, 0.962]$ & 0.759 & 0.959 & 0.006 \\
      \quad RV enlargement           & 0.882 $[0.833, 0.930]$ & 0.738 & 0.837 & 0.009 \\
      \quad LV enlargement           & 0.868 $[0.843, 0.895]$ & 0.785 & 0.811 & 0.038 \\
      \quad Mitral stenosis          & 0.836 $[0.685, 0.971]$ & 0.556 & 0.956 & 0.002 \\
      \quad Aortic stenosis          & 0.835 $[0.734, 0.917]$ & 0.444 & 0.970 & 0.004 \\
      \quad Mitral regurgitation     & 0.808 $[0.739, 0.870]$ & 0.707 & 0.814 & 0.009 \\
      \quad Pericardial effusion     & 0.728 $[0.676, 0.777]$ & 0.785 & 0.540 & 0.021 \\
      \quad Ascending aortic dil.    & 0.685 $[0.658, 0.713]$ & 0.640 & 0.643 & 0.073 \\
      \quad LA enlargement           & 0.665 $[0.648, 0.680]$ & 0.740 & 0.510 & 0.201 \\
      \midrule
      \multicolumn{5}{l}{\textbf{Temporal Test Set from PKUPH}} \\
      \quad Global heart enlargement & 0.929 $[0.883, 0.963]$ & 0.730 & 0.930 & 0.006 \\
      \quad Reduced LV systolic fn.  & 0.920 $[0.901, 0.936]$ & 0.893 & 0.767 & 0.023 \\
      \quad RA enlargement           & 0.913 $[0.887, 0.936]$ & 0.863 & 0.824 & 0.014 \\
      \quad Tricuspid regurgitation  & 0.893 $[0.837, 0.936]$ & 0.733 & 0.876 & 0.008 \\
      \quad Mitral stenosis          & 0.889 $[0.735, 0.977]$ & 0.636 & 0.952 & 0.002 \\
      \quad Mitral regurgitation     & 0.842 $[0.786, 0.889]$ & 0.600 & 0.895 & 0.009 \\
      \quad LV enlargement           & 0.838 $[0.812, 0.861]$ & 0.741 & 0.767 & 0.041 \\
      \quad Aortic regurgitation     & 0.807 $[0.721, 0.886]$ & 0.517 & 0.911 & 0.005 \\
      \quad RV enlargement           & 0.771 $[0.689, 0.846]$ & 0.544 & 0.888 & 0.009 \\
      \quad Aortic stenosis          & 0.756 $[0.616, 0.883]$ & 0.214 & 0.972 & 0.003 \\
      \quad Pericardial effusion     & 0.691 $[0.646, 0.734]$ & 0.543 & 0.764 & 0.028 \\
      \quad Ascending aortic dil.    & 0.677 $[0.657, 0.700]$ & 0.633 & 0.629 & 0.089 \\
      \quad LA enlargement           & 0.657 $[0.641, 0.673]$ & 0.543 & 0.683 & 0.203 \\
      \midrule
      \multicolumn{5}{l}{\textbf{External Test Set from SHTMU}} \\
      \quad Mitral stenosis          & 0.906 $[0.837, 0.951]$ & 0.682 & 0.922 & 0.001 \\
      \quad Global heart enlargement & 0.877 $[0.860, 0.894]$ & 0.653 & 0.883 & 0.018 \\
      \quad Reduced LV systolic fn.  & 0.866 $[0.858, 0.874]$ & 0.803 & 0.776 & 0.074 \\
      \quad Pericardial effusion     & 0.832 $[0.790, 0.870]$ & 0.811 & 0.724 & 0.009 \\
      \quad Tricuspid regurgitation  & 0.820 $[0.800, 0.837]$ & 0.664 & 0.833 & 0.031 \\
      \quad Mitral regurgitation     & 0.806 $[0.788, 0.823]$ & 0.783 & 0.725 & 0.031 \\
      \quad RA enlargement           & 0.803 $[0.762, 0.843]$ & 0.732 & 0.744 & 0.008 \\
      \quad RV enlargement           & 0.748 $[0.670, 0.819]$ & 0.622 & 0.776 & 0.005 \\
      \quad Aortic stenosis          & 0.714 $[0.576, 0.832]$ & 0.182 & 0.940 & 0.001 \\
      \quad LV enlargement           & 0.698 $[0.605, 0.777]$ & 0.815 & 0.506 & 0.022 \\
      \quad Aortic regurgitation     & 0.639 $[0.609, 0.669]$ & 0.343 & 0.805 & 0.015 \\
      \quad Ascending aortic dil.    & 0.635 $[0.622, 0.648]$ & 0.648 & 0.545 & 0.094 \\
      \quad LA enlargement           & 0.610 $[0.602, 0.618]$ & 0.617 & 0.548 & 0.283 \\
      \bottomrule
    \end{tabular}
    \begin{tablenotes}
      \footnotesize
      \item LV, left ventricular; LA, left atrial; RV, right ventricular; RA, right atrial; dil., dilation; fn., function.
    \end{tablenotes}
  \end{threeparttable}
\end{table}

\subsection*{Methodological and per-label performance comparison with contemporary AI-ECG models}

To contextualize both the methodological contributions and diagnostic performance of AnyECG-Echo, we conducted a systematic comparison against state-of-the-art AI-ECG models for SHD detection published between 2025 and 2026 (Table~\ref{tab:per_label_comparison}). The upper panel summarizes each model across five methodological dimensions; the lower panel presents per-label AUROC values on each model's respective test cohort. As these comparisons are cross-cohort rather than head-to-head, direct numerical comparison should be interpreted with caution.

A central finding of this comparison is the data efficiency of AnyECG-Echo. With only 37,389 ECG--echocardiography pairs—5 to 33 times fewer than its comparators—AnyECG-Echo achieves competitive or superior discrimination across the conditions where direct comparison is possible. For reduced LV systolic function, AnyECG-Echo attains an internal AUROC of 0.924 using single-lead input, compared with 0.904 for the 12-lead EchoNext trained on over 1.2 million pairs.

Equally important is the diagnostic breadth that this efficiency enables. While EchoNext and EchoNext-Mini cover 11--12 categories and Wearable-Echo-FM targets only 3, AnyECG-Echo resolves 13 fine-grained subtypes from a single-lead waveform. Seven of these—including global heart enlargement, chamber-specific enlargements (LA, LV, RA, RV), ascending aortic dilation, and mitral stenosis—have no published AI-ECG comparator to date, representing unique diagnostic coverage.

Beyond performance, the comparison reveals complementary gaps in existing approaches. EchoNext~\cite{poterucha2025detecting} and AI-ECG VHD~\cite{liang2025artificial} demonstrate strong multi-centre performance but remain dependent on clinical 12-lead ECG acquisition, limiting their deployment in home-monitoring or point-of-care settings. Wearable-Echo-FM~\cite{Wearable-Echo-FM} pioneered single-lead contrastive learning but was validated only within a single health system and without mechanistic interpretability. AnyECG-Echo is, to our knowledge, the only framework that simultaneously achieves native single-lead compatibility, geographically independent external validation, fine-grained multi-subtype coverage, and dual-axis physiological interpretability—demonstrating that a small, well-structured dataset paired with report-supervised contrastive learning can match or exceed the diagnostic utility of models trained on orders-of-magnitude more data.

\begin{table*}[t]
\centering
\caption{Per-label AUROC comparison between AnyECG-Echo and contemporary AI-ECG models, with summary of methodological characteristics. For AnyECG-Echo, the internal cohort corresponds to the PKUPH internal test, and the external cohort corresponds to the SHTMU external validation. All AUROC values represent performance on each model's respective test cohort. Dashes indicate that the condition was not evaluated by the corresponding model.}
\label{tab:per_label_comparison}
\renewcommand{\arraystretch}{1.25}
\resizebox{\textwidth}{!}{%
\begin{tabular}{@{} l cc c c cc c @{}}
\toprule
 & \multicolumn{2}{c}{\textbf{AnyECG-Echo (Ours)}}
 & \textbf{EchoNext~\cite{poterucha2025detecting}}
 & \textbf{EchoNext-Mini~\cite{hughes2026echonext}}
 & \multicolumn{2}{c}{\textbf{AI-ECG VHD~\cite{liang2025artificial}}}
 & \textbf{Wearable-Echo-FM~\cite{Wearable-Echo-FM}} \\
\cmidrule(lr){2-3} \cmidrule(lr){4-4} \cmidrule(lr){5-5} \cmidrule(lr){6-7} \cmidrule(lr){8-8}
 & \textbf{Internal} & \textbf{External}
 & {\footnotesize Multi-centre}
 & {\footnotesize Multi-centre}
 & {\footnotesize Internal} & {\footnotesize External}
 & {\footnotesize Internal} \\
\midrule
\multicolumn{8}{@{}l}{\textit{Model characteristics}} \\[2pt]
\quad Input type
 & \multicolumn{2}{c}{1-lead}
 & 12-lead
 & 12-lead
 & \multicolumn{2}{c}{12-lead}
 & 1-lead \\
\quad Dataset scale
 & \multicolumn{2}{c}{37,389 pairs}
 & 1,245,273 pairs
 & 100,000 pairs
 & \multicolumn{2}{c}{400,882 pairs}
 & 194,551 pairs \\
\quad Diagnostic granularity
 & \multicolumn{2}{c}{13 subtypes}
 & 11 categories
 & 12 categories
 & \multicolumn{2}{c}{3 categories}
 & 3 categories \\
\quad External validation
 & \multicolumn{2}{c}{\checkmark}
 & \checkmark
 & ---
 & \multicolumn{2}{c}{\checkmark}
 & \texttimes \\
\quad Interpretability
 & \multicolumn{2}{c}{\checkmark}
 & \texttimes
 & \texttimes
 & \multicolumn{2}{c}{\checkmark}
 & \texttimes \\
\quad Source
 & \multicolumn{2}{c}{-}
 & \textit{Nature} 2025
 & \textit{NEJM AI} 2026
 & \multicolumn{2}{c}{\textit{NEJM AI} 2025}
 & \textit{EHJ DH} 2026 \\
\midrule
\multicolumn{8}{@{}l}{\textit{Per-label AUROC}} \\[2pt]
\quad Mitral stenosis & 0.836[0.685, 0.971] & 0.906[0.837, 0.951] & --- & --- & --- & --- & --- \\
\quad Global heart enlargement & 0.931[0.880, 0.971] & 0.877[0.860, 0.894] & --- & --- & --- & --- & --- \\
\quad Reduced LV systolic function & 0.924[0.895, 0.950] & 0.866[0.858, 0.874] & 0.904[0.896, 0.912] & 0.852[0.839, 0.865] & --- & --- & 0.894[0.884, 0.903] \\
\quad Pericardial effusion & 0.728[0.676, 0.777] & 0.832[0.790, 0.870] & 0.799[0.761, 0.834]& 0.766[0.712, 0.818] & --- & --- & --- \\
\quad Tricuspid regurgitation & 0.907[0.838, 0.962] & 0.820[0.800, 0.837] & --- & 0.833[0.812, 0.852] & 0.910[0.905, 0.914] & 0.805[0.795, 0.816] & --- \\
\quad Mitral regurgitation& 0.808[0.739, 0.870] & 0.806[0.788, 0.823] & 0.855[0.841, 0.869] & 0.806[0.782, 0.829] & 0.893[0.889, 0.897] & 0.777[0.767, 0.785] & --- \\
\quad RA enlargement & 0.914[0.871, 0.952] & 0.803[0.762, 0.843] & --- & --- & --- & --- & --- \\
\quad RV enlargement & 0.882[0.833, 0.930] & 0.748[0.670, 0.819] & --- & --- & --- & --- & --- \\
\quad Aortic stenosis & 0.835[0.734, 0.917] & 0.714[0.576, 0.832] & 0.864[0.850, 0.877] & 0.859[0.839, 0.878] & --- & --- & --- \\
\quad LV enlargement & 0.868[0.843, 0.895] & 0.698[0.605, 0.777] & --- & --- & --- & --- & --- \\
\quad Aortic regurgitation & 0.928[0.841, 0.979] & 0.639[0.609, 0.669] & 0.777[0.749, 0.805] & 0.740[0.688, 0.794] & 0.845[0.838, 0.851] & 0.751[0.731, 0.774] & --- \\
\quad Ascending aortic dilation & 0.685[0.658, 0.713] & 0.635[0.622, 0.648] & --- & --- & --- & --- & --- \\
\quad LA enlargement & 0.665[0.648, 0.680] & 0.610[0.602, 0.618] & --- & --- & --- & --- & --- \\
\bottomrule
\multicolumn{8}{@{}l}{\footnotesize \textit{Note}: Internal and External validation cohorts vary by model study definition (e.g., EchoNext's ``Multi-centre'' refers to its respective multi-center test dataset).} \\
\end{tabular}%
}
\end{table*}

\subsection*{Demographic subgroup analysis and age-related performance attenuation}

We evaluated model robustness across age and sex strata in the PKUPH cohort (Table~\ref{tab:renming_subgroup_template}). Sex-stratified analysis revealed consistent performance between male and female patients, with overlapping confidence intervals confirming an absence of systematic gender-related bias. In contrast, model performance exhibited a monotonic decline across most categories as age increased from $<50$ to $\ge 75$ years. This attenuation likely stems from age-associated electrophysiological confounders, such as conduction delays and atrial fibrillation, which accumulate in older populations and obscure subtle structural signatures. 

\begin{table*}[t]
\centering
\caption{\textbf{Subgroup analysis of model performance in the external cohort}, stratified by age group and sex. Values are reported as AUROC [95\% confidence interval].}
\label{tab:renming_subgroup_template}
\setlength{\tabcolsep}{8pt}
\renewcommand{\arraystretch}{1.35}
\resizebox{\textwidth}{!}{%
\begin{tabular}{@{} l c c c c c c @{}}
\toprule

\textbf{Subgroup}
    & \makecell{\textbf{SHD ($n$)}\\(pos/neg)}
    & \makecell{\textbf{Reduced LV}\\\textbf{systolic function}}
    & \makecell{\textbf{LV}\\\textbf{enlargement}}
    & \makecell{\textbf{Pericardial}\\\textbf{effusion}}
    & \makecell{\textbf{Ascending aortic}\\\textbf{dilation}}
    & \makecell{\textbf{SHD}\\\textbf{(overall)}} \\

\midrule

\rowcolor[gray]{0.95}
\multicolumn{7}{@{}l}{\textbf{Age group (years)}} \\

$<$50    & 142/280
  & 0.912 [0.807, 0.987]
  & 0.841 [0.753, 0.919]
  & 0.773 [0.662, 0.875]
  & 0.823 [0.686, 0.934]
  & 0.683 [0.628, 0.736] \\

50--64   & 347/410
  & 0.939 [0.909, 0.964]
  & 0.892 [0.839, 0.935]
  & 0.845 [0.746, 0.931]
  & 0.617 [0.542, 0.690]
  & 0.691 [0.652, 0.729] \\

65--74   & 342/248
  & 0.961 [0.934, 0.982]
  & 0.790 [0.702, 0.871]
  & 0.773 [0.593, 0.922]
  & 0.613 [0.535, 0.691]
  & 0.641 [0.596, 0.683] \\

$\geq$75   & 267/136
  & 0.855 [0.685, 0.978]
  & 0.860 [0.783, 0.922]
  & 0.621 [0.449, 0.795]
  & 0.635 [0.566, 0.707]
  & 0.677 [0.622, 0.730] \\

\midrule

\rowcolor[gray]{0.95}
\multicolumn{7}{@{}l}{\textbf{Sex}} \\

Male      & 858/571
  & 0.921 [0.886, 0.952]
  & 0.862 [0.830, 0.895]
  & 0.773 [0.681, 0.859]
  & 0.655 [0.612, 0.698]
  & 0.708 [0.681, 0.733] \\

Female     & 753/999
  & 0.920 [0.871, 0.961]
  & 0.817 [0.749, 0.878]
  & 0.708 [0.636, 0.772]
  & 0.685 [0.634, 0.733]
  & 0.697 [0.672, 0.721] \\

\bottomrule
\multicolumn{7}{@{}p{1.48\textwidth}}{\footnotesize
  \textit{n} = positive / negative ECG recordings per subgroup (SHD overall). SHD, structural heart disease. LV, Left ventricular.}
\end{tabular}%
}
\end{table*}


\subsection*{AnyECG-Echo modeling achieves consistency with canonical medical interpretations}

To verify that AnyECG-Echo prioritizes clinically valid signals over center-specific shortcuts, we performed a hierarchical interpretability analysis that bridges population-level trends and individual-level evidence. We first compared population-level median beats between high-risk and low-risk groups to capture macroscopic morphological shifts. This population perspective was then cross-validated at the individual level using occlusion-based Shapley value attribution, specifically targeting the patients identified by the model as having the highest diagnostic risk. By integrating these two perspectives, we can evaluate whether the general pathological signatures learned by the model are consistently utilized to drive its most confident clinical predictions.

\begin{figure*}[t]
    \centering
    \includegraphics[width=1.0\textwidth]{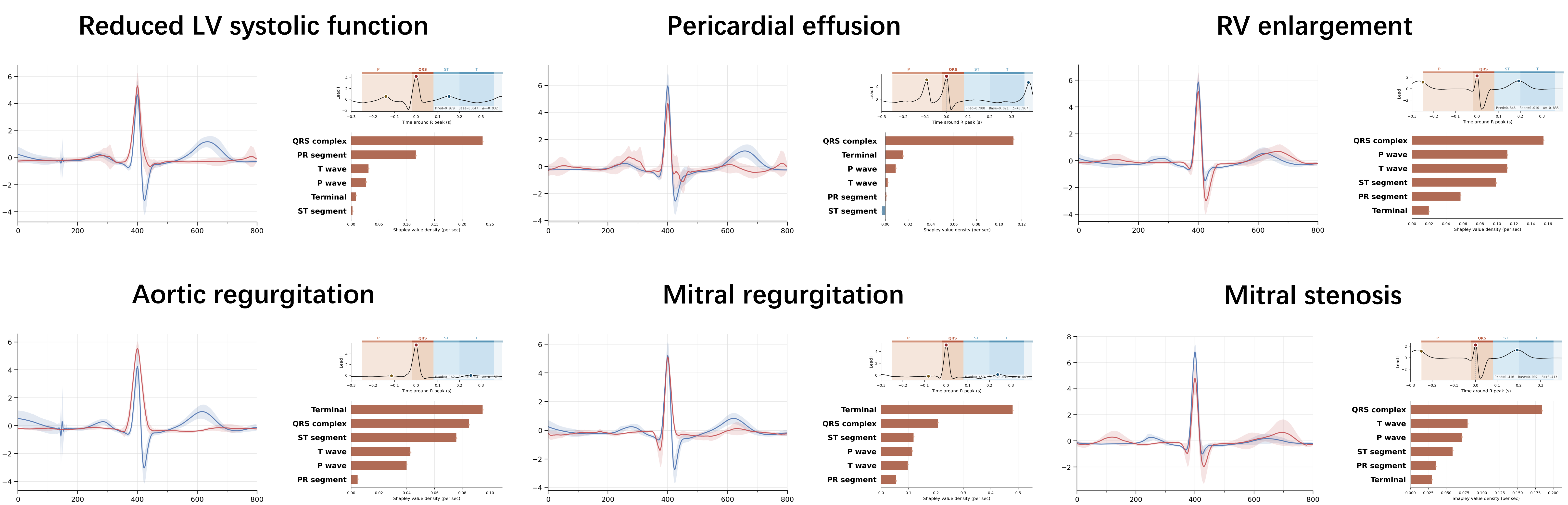}
\caption{\textbf{Electrophysiological interpretation of learned ECG features across SHD phenotypes.} Each cardiac condition is visualized as a two-panel vertical composite designed to bridge population-level trends and individual-level evidence. The left panels display population-level median beats for high-risk (red) versus low-risk (blue) groups, illustrating macroscopic morphological shifts. The right panels present individual-level segment attribution maps (Shapley values) for patients identified by the model as having the highest diagnostic risk, highlighting the specific features driving its most confident predictions. Attribution was computed via occlusion sensitivity: for each of six predefined ECG segments, all corresponding intervals were replaced with zero values (equivalent to the population-mean amplitude after z-score normalization), and the resulting decrease in predicted probability was normalized by masked duration to yield a Shapley value density (per second). Scores were averaged across the top-5 highest-probability patients per condition (minimum threshold 0.5); individual patient-level values are overlaid as scatter points to illustrate inter-patient consistency.}
    \label{fig:morphology}
\end{figure*}

Specifically, six physiologically defined ECG segments were delineated using fixed temporal offsets relative to each detected R-peak: P wave ($-$250 to $-$100 ms), PR segment ($-$100 to $-$20 ms), QRS complex ($-$20 to +80 ms), ST segment (+80 to +200 ms), T wave (+200 to +360 ms), and terminal segment (+360 to +400 ms). For each segment, all corresponding intervals across every detected heartbeat in a given recording were simultaneously replaced with zero values. Because the input signal undergoes z-score normalization prior to inference, zero-fill replacement is equivalent to substituting the population-mean amplitude, serving as a physiologically neutral baseline. The attribution score for each segment was then defined as the decrease in predicted probability divided by the total masked duration in seconds (termed Shapley value density), enabling fair comparison across segments of different lengths. For each SHD subtype, this analysis was performed on the top-5 patients ranked by predicted probability (with a minimum threshold of 0.5), and scores were averaged across this cohort; individual patient-level values are shown as scatter points in Fig.~\ref{fig:morphology} to demonstrate inter-patient consistency. Population-level median beats were extracted by segmenting individual heartbeats within 800 ms windows centered on R-peaks, excluding morphological outliers (Pearson $r < 0.8$ relative to the initial median template), and computing the element-wise median of the remaining waveforms.

The primary mechanistic evidence is presented in Fig.~\ref{fig:morphology}. AnyECG-Echo effectively captures diagnostic signatures across the QRS complex, ST segment, and T wave, demonstrating high consistency between population-level morphological shifts and individual-level attributions.

For \textbf{Reduced LV systolic function}, although no direct canonical ECG correlate exists, our findings align with established explainability results for $LVEF$~\cite{van2025explainable}, with both median beats and Shapley values primarily localized to the $R$-wave. For \textbf{pericardial effusion}, we observed diminished QRS amplitude in median beats, with the highest Shapley attribution concentrated on the QRS complex. \textbf{RV enlargement} was characterized by deepened $S$-waves and reduced $R$-waves, with model attribution again localized to the QRS complex.

For valvular conditions, \textbf{aortic regurgitation} median beats exhibited increased $R$-wave voltage alongside $ST$-segment depression and $T$-wave inversion; accordingly, Shapley values were elevated across the QRS, $ST$, and $T$ segments. \textbf{Mitral regurgitation} showed similar $ST$-$T$ perturbations, with substantial attribution distributed across the QRS and $ST$-$T$ segments. Finally, for \textbf{mitral stenosis}, median beats displayed widened $P$-waves, reduced $R$-waves, and deepened $S$-waves, with model attribution correctly prioritizing the $P$-wave and QRS complex.

Collectively, these findings demonstrate that AnyECG-Echo achieves fundamental consistency between deep-learning modeling and established medical interpretation. By grounding its predictions in canonical electrophysiology at both the population and individual scales, the framework ensures that its structural insights are mechanistically valid, transparent, and clinically trustworthy.

\subsection*{Predictive probabilities from AnyECG-Echo function as quantitative digital biomarkers}

We calculated Spearman rank correlations between predicted disease probabilities and three gold-standard metrics: left ventricular ejection fraction ($LVEF$), interventricular septum thickness ($IVSs$), and LV posterior wall thickness ($LVPWS$). 

\begin{figure*}[h]  
    \centering  
    \includegraphics[width=1.0\textwidth]{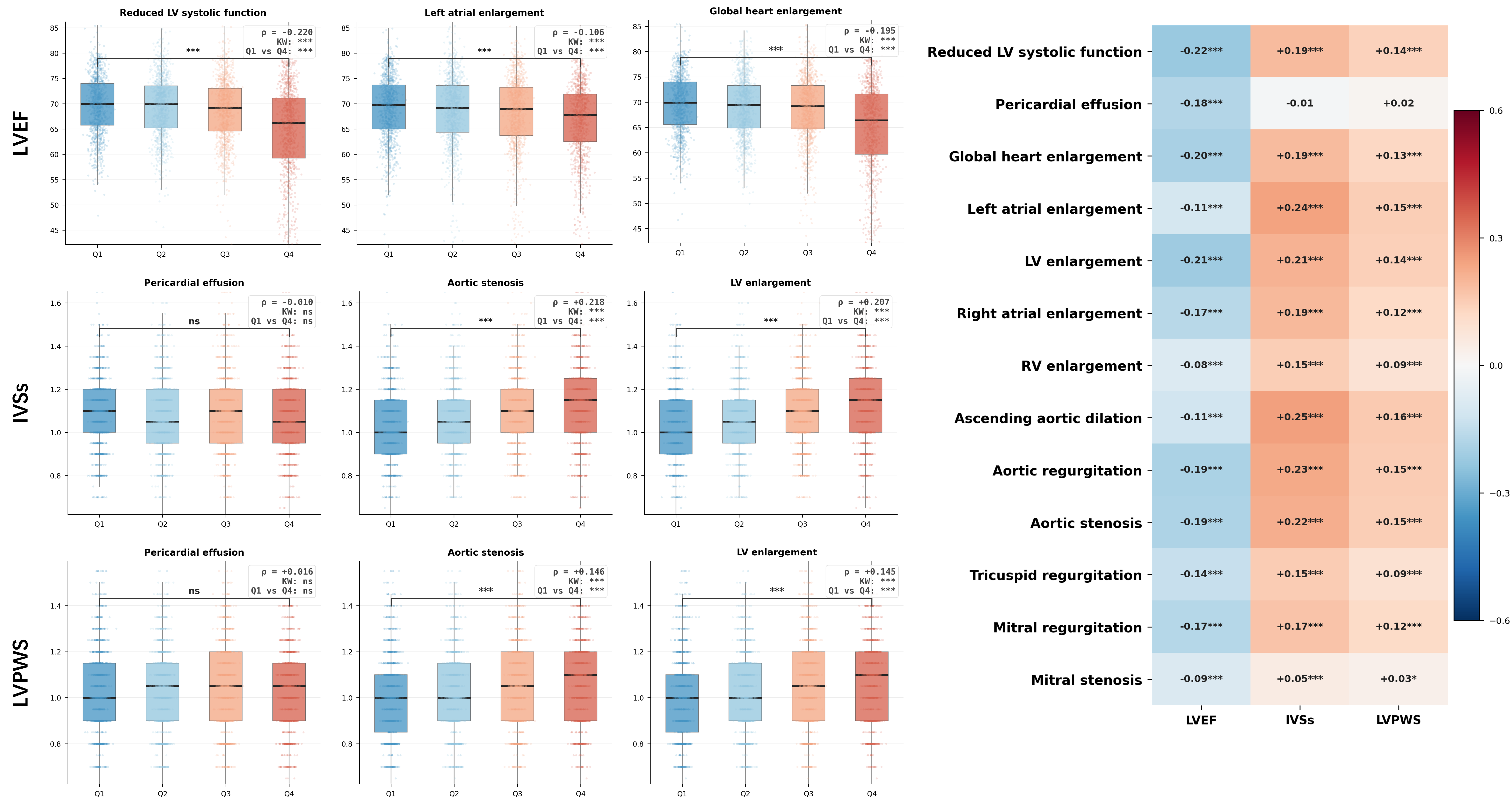}  
    \caption{\textbf{Spearman correlations between AnyECG-Echo predictions and quantitative echocardiographic parameters.} Left: representative boxplots showing echo parameter values stratified by predicted probability quartile (Q1--Q4). Right: heatmap of Spearman $\rho$ across all subtypes and parameters.}
    \label{fig:correlation}  
\end{figure*}

As shown in Fig.~\ref{fig:correlation}, predicted probabilities for reduced LV systolic function and global heart enlargement demonstrated robust negative correlations with $LVEF$ ($\rho = -0.220$ and $-0.195$, respectively; all $p < 0.001$). For pressure-overload conditions such as aortic stenosis and LV enlargement, predicted probabilities accurately tracked increases in wall thickness ($IVSs$ and $LVPWS$). 

Crucially, for these examples, we observed in the boxplots on the left that the predicted probability quartiles (Q1--Q4) maintain a monotonic correspondence with the echocardiographic parameters across different intervals. This confirms that AnyECG-Echo tracks the graded severity of systolic impairment, allowing the single-lead waveform to serve as a continuous digital biomarker of cardiac output.

We also identified a near-zero correlation with myocardial thickness ($IVSs$: $\rho = -0.01$; $LVPWS$: $\rho = +0.02$). This selective sensitivity ensures that the model discriminates between distinct disease mechanisms—fluid accumulation versus structural remodeling—rather than relying on non-specific signals of abnormality.

These findings establish AnyECG-Echo as a physiologically grounded digital biomarker that transcends binary diagnostic categories. This capability enables continuous, data-driven risk stratification in community and home-monitoring settings, identifying clinically significant structural decline long before the onset of symptomatic decompensation.

\subsection*{Data efficiency and asymptotic performance in downstream SHD detection}

To evaluate the scaling laws of AnyECG-Echo, we fine-tuned the model on varying subsets of the development cohort (total patients $N=12,726$), ranging from ultra-low sample sizes (100, 500, and 1,273 patients) to larger fractions (30\%, 50\%, 80\%, and 100\%) of the total dataset.

\begin{figure*}[h]
\centering
\includegraphics[width=1.0\textwidth]{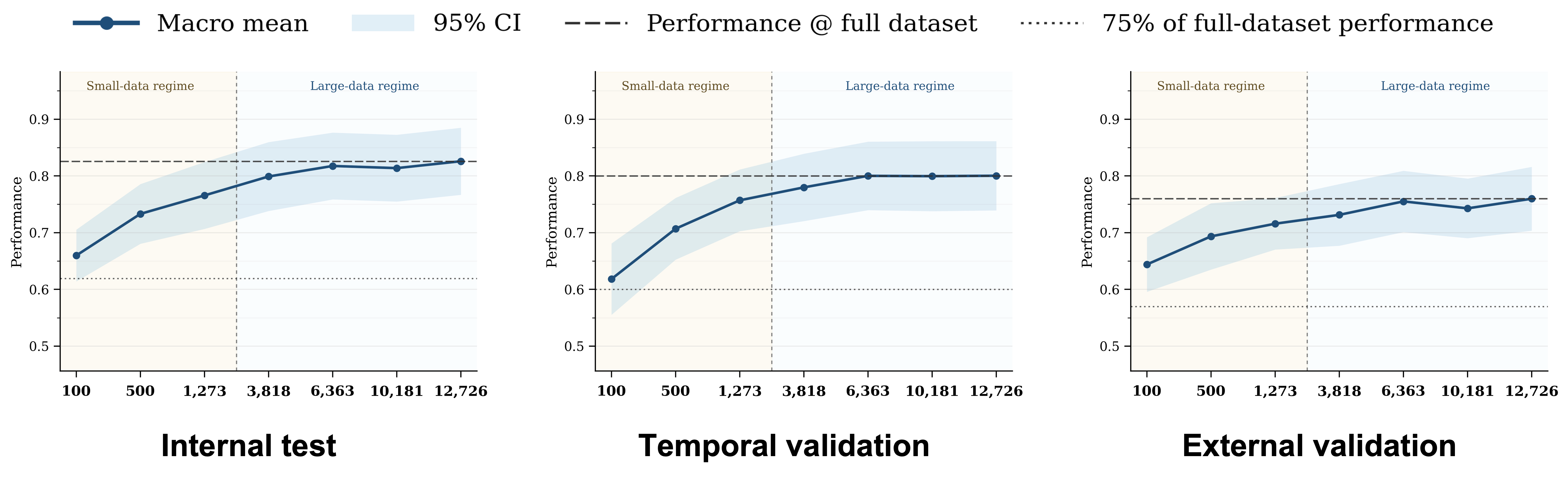}
\caption{\textbf{Data efficiency and scaling laws for downstream SHD detection.} 
Scaling of model performance (Macro mean AUROC) across the internal test, temporal validation, and external validation cohorts as a function of training sample size. 
In the small-data regime (yellow), AnyECG-Echo exhibits a rapid gain in discriminative ability, with only 100 labeled samples exceeding 75\% of the full-dataset benchmark (dotted line). 
In the large-data regime (blue), performance follows an asymptotic trajectory, essentially reaching its diagnostic ceiling at 50\% of the total dataset (dashed line). 
Shaded areas represent 95\% confidence intervals.}
\label{fig:scaling}
\end{figure*}

As shown in Fig.~\ref{fig:scaling}, AnyECG-Echo exhibited remarkable data efficiency across all cohorts. In the small-data regime, we observed a sharp and rapid increase in discriminative performance. Notably, with an ultra-minimal budget of only 100 labeled samples, the model’s performance already exceeded 75\% of the level achieved by the full dataset. 

As the sample size increased into the large-data regime, the performance followed a clear asymptotic trajectory. By the 50\% sampling mark (representing 6,363 patients), AnyECG-Echo essentially reached its diagnostic ceiling, matching the performance attained by the full-dataset configuration. These results demonstrate that the multimodal pre-training successfully internalized a high-quality representation of cardiac physiology, enabling robust downstream adaptation with minimal supervised fine-tuning. This property is critical for deploying the framework in specialized clinical settings where labeled echocardiographic data are scarce.

\subsection*{AnyECG-Echo provides net clinical benefit across the structural heart disease spectrum}

\begin{figure*}[h]
\centering
\includegraphics[width=1.0\textwidth]{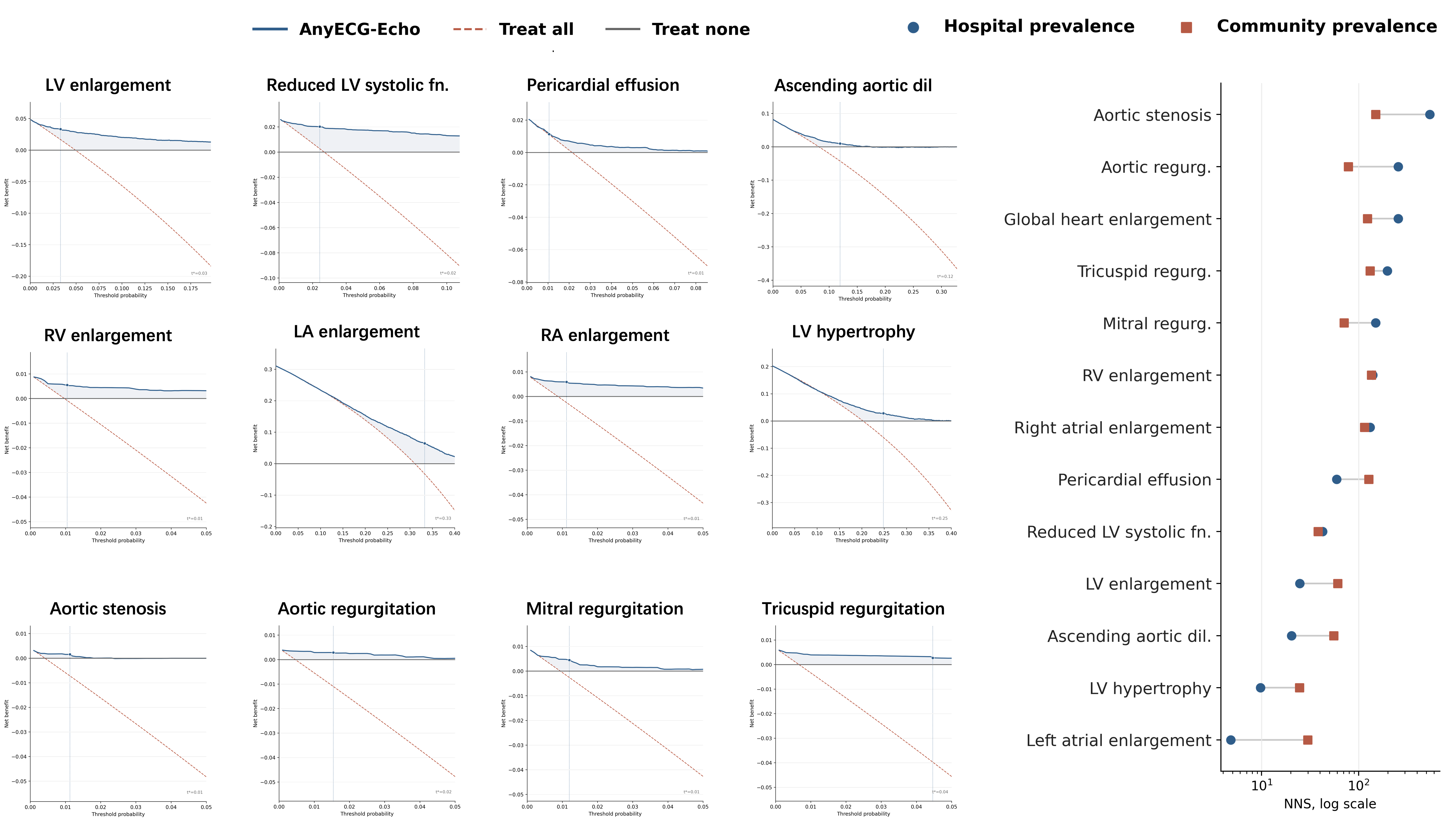}
\caption{\textbf{Clinical utility of AnyECG-Echo for SHD screening.} Left panels show decision curve analyses (net benefit versus threshold probability) for each of the 13 SHD subtypes evaluated on the internal test cohort. The blue line represents AnyECG-Echo, the red dashed line represents the strategy of referring all patients for echocardiography (treat all), and the grey line represents the strategy of referring none (treat none). AnyECG-Echo provides positive net benefit above both reference strategies across the range of clinically relevant thresholds. The right panel displays screening efficiency (number of ECG screens required to detect one true-positive case, log scale) under hospital prevalence (blue circles) and literature-derived community prevalence (orange squares).}
\label{fig:dca_nns}
\end{figure*}

To evaluate the clinical utility of AnyECG-Echo beyond discrimination metrics, we performed decision curve analysis and estimated screening efficiency for 12 SHD subtypes on the internal test cohort (Fig.~\ref{fig:dca_nns}). Across the full range of clinically plausible threshold probabilities, AnyECG-Echo consistently yielded a higher net benefit than both the default strategies of screening all patients and screening none, confirming that model-guided referral for echocardiography reduces unnecessary testing without sacrificing case detection.

Screening efficiency, quantified as the number of single-lead ECGs required to identify one true-positive case, further supports practical deployment. Under hospital prevalence, AnyECG-Echo required fewer than 10 screens per detection for LV enlargement, and Global heart enlargement. When recalculated under literature-derived community prevalence estimates to simulate a population-level screening scenario, the number of screens per detection increased as expected but remained within a practically actionable range (median across most subtypes $<$ 100). Given that a single-lead ECG can be acquired in under 30 seconds at negligible cost using a consumer-grade wearable device, these figures compare favourably with established screening programmes for other asymptomatic conditions and support the feasibility of large-scale, opportunistic SHD screening outside the hospital setting.

\section*{DISCUSSION}
Our results demonstrate that AnyECG-Echo enables the identification of complex structural heart disease from single-lead ECGs by systematically aligning electrophysiological waveforms with echocardiographic evidence. By leveraging multimodal contrastive learning, this study establishes a scalable approach to extract structural diagnostic information from single-lead signals, facilitating a screening solution compatible with population-level cardiac monitoring. This framework provides fine-grained diagnosis across 13 distinct SHD subtypes. The clinical utility of these predictions is underpinned by two objective properties: physiological consistency and functional performance as a quantitative digital biomarker.

Physiological consistency is established through the alignment of population-level morphological variations with individual-level feature attributions. Analysis of population-level median heartbeats reveals distinct macro-morphological shifts between high-risk and low-risk groups, while patient-specific Shapley-value maps further localize the exact electrophysiological features driving the model’s most confident diagnostic predictions. This dual-level validation confirms that the model’s internal representations are grounded in established cardiac pathophysiology rather than non-specific statistical correlations.

The framework further operates as a quantitative digital biomarker, as evidenced by significant Spearman correlations between model predictions and gold-standard echocardiographic parameters. These correlations demonstrate that the continuous predictive probabilities can serve as physiological proxies for structural measurements, allowing for a graded assessment of disease severity. By providing a continuous risk metric that tracks quantitative cardiac remodeling, the system supports risk-proportional clinical decision-making beyond simple binary classification.

From a public health perspective, AnyECG-Echo utilizes routinely generated, natural-language echocardiography reports as a supervisory signal. This approach ensures the model remains robust against geographic and temporal shifts, maintaining high discriminative performance across external validation cohorts. Such scalability facilitates an optimized diagnostic pathway where single-lead ECGs serve as a preliminary screening tool to improve clinical resource allocation alongside definitive imaging modalities.

While this study validates AnyECG-Echo using clinical Lead~I recordings in hospital settings, the framework's single-lead architecture and lightweight inference suggest several potential deployment scenarios that warrant future investigation:

\begin{itemize}
    \item \textbf{Opportunistic screening in primary care}: Standard ECGs acquired during routine visits could be automatically screened for SHD risk, flagging high-probability cases for specialist referral. Prospective validation in primary care populations with lower disease prevalence is needed to establish real-world screening performance.
    \item \textbf{Wearable-based home monitoring}: Consumer-grade single-lead devices could enable longitudinal risk tracking to detect progressive structural changes over time. However, clinical Lead~I serves only as a proxy for wearable signals, and dedicated validation on device-specific recordings with motion artifacts and variable signal quality remains a prerequisite.
    \item \textbf{Emergency triage}: Rapid structural assessment from a single-lead waveform could inform pre-hospital resource allocation. This application requires prospective evaluation against established triage protocols before clinical adoption.
\end{itemize}

Several limitations of this study warrant consideration. The primary development and validation in cohorts of East Asian descent necessitate further cross-ethnic validation in Western populations to ensure universal generalizability. While temporal and geographic validations were utilized to mitigate overfitting, the retrospective nature of this work requires future prospective, real-world trials to evaluate the model's true impact on clinical workflows and patient outcomes.

The quality of the supervisory signal remains dependent on the diagnostic consistency of physician-authored echocardiography reports. Although contrastive learning is robust to noise, the model's sensitivity to interval cardiac changes within the $\pm$10-day temporal matching window has not been specifically evaluated through ablation studies. Furthermore, data completeness remains a challenge; for instance, the 41.5\% missing sex data in specific cohort subsets may limit the robustness and interpretability of sex-stratified bias analyses.

Anatomical and hardware-specific constraints also influence the model's scope. AnyECG-Echo demonstrated limited discriminative power for atrial pathologies, likely due to the single-lead configuration's inherent difficulty in capturing subtle $P$-wave morphologies compared to the more dominant ventricular signals of the QRS complex. Additionally, clinical Lead~I recordings serve only as a proxy for wearable devices and lack the motion artifacts and longitudinal dynamics characteristic of real-world consumer hardware. Future research must bridge the gap between clinical-grade data and noisy, continuous recordings from diverse hardware platforms to enable true home-based monitoring.

Spectrum bias inherent in the study populations also influences the interpreted performance. Both cohorts comprised patients clinically referred for echocardiography rather than general community populations, resulting in a dataset enriched for symptomatic and advanced structural heart disease. The reported AUROC values consequently reflect performance within a referral-enriched setting, meaning that community number-needed-to-screen estimates have not been validated against actual screening outcomes, despite being calculated using literature-derived prevalence figures. Prospective, community-based cohorts remain necessary to establish real-world performance in asymptomatic populations with lower baseline disease prevalence.

The temporal validation set, partitioned via a November 2019 cutoff, functions as an internal temporal hold-out rather than a fully independent temporal validation. Because this subset shares the same institutional infrastructure, hardware platforms, and physician workforce as the development cohort, it does not capture the complex distributional shifts associated with long-term changes in clinical practice, equipment upgrades, or staff turnover. Although the geographically independent external cohort from SHTMU partially mitigates this specific institutional bias, multi-center, prospectively enrolled temporal validation across diverse clinical environments is required to demonstrate definitive long-term robustness.

\section*{METHODS}

\subsection*{Definitions of 13 SHD subtype labels}

Labels were derived from physician-authored echocardiography reports. Each label was assigned as positive when the corresponding echocardiographic finding met or exceeded the threshold criteria defined below. Definitions follow current American Society of Echocardiography (ASE) where applicable~\cite{lang2015recommendations}.
Each valvular label denotes moderate or greater severity, defined by the qualitative or semi-quantitative thresholds above, derived from the integrative grading scheme of the 2017 ASE/SCMR (regurgitation) and 2017 EACVI/ASE (stenosis) guidelines
~\cite{zoghbi2017recommendations, baumgartner2017recommendations}. As labels were extracted by rule-based nature language process (NLP) from physician-authored reports, the dominant signal is the reported severity term; the numeric thresholds are listed for transparency rather than as the literal extraction rule.

\begin{table*}[h]
\centering
\caption{Definitions and diagnostic criteria for the 13 SHD subtype labels.}
\label{tab:label_definitions}
\renewcommand{\arraystretch}{1.35}
\resizebox{\textwidth}{!}{%
\begin{tabular}{@{} l l @{}}
\toprule
\textbf{Label} & \textbf{Echocardiographic criteria} \\
\midrule
Reduced LV systolic function & LVEF $<$ 53\% \\
Pericardial effusion & Echo-free space in diastole $>$ 10 mm \\
LV enlargement & LVIDd $>$ 5.8 cm (male) or $>$ 5.2 cm (female); or LV EDV/BSA $>$ 74 mL/m$^2$ (male) or $>$ 61 mL/m$^2$ (female) \\
LA enlargement & LAD $>$ 4.0 cm (male) or $>$ 3.8 cm (female); or LA volume index $>$ 34 mL/m$^2$ \\
RV enlargement & RVD1 $>$ 41 mm at basal level; or RVD2 $>$ 35 mm at mid-level \\
RA enlargement & RA volume index $>$ 39 mL/m$^2$ (male) or $>$ 33 mL/m$^2$ (female); or RA minor axis $>$ 2.5 cm/m$^2$ \\
Global heart enlargement & Enlargement of $\geq$3 chambers reported simultaneously \\
Aortic stenosis (mod+)
& Mean gradient $\geq$ 20 mmHg, or peak aortic jet velocity ($V_{\max}$) $\geq$ 3.0 m/s, or AVA $\leq$ 1.5 cm$^2$ (indexed AVA $\leq$ 0.85 cm$^2$/m$^2$) \\
Aortic regurgitation (mod+)
& VC $\geq$ 3 mm, or EROA $\geq$ 0.10 cm$^2$, or RVol $\geq$ 30 mL, or RF $\geq$ 30\%, or PHT $\leq$ 500 ms, or any holodiastolic flow reversal in the descending aorta \\
Mitral stenosis (mod+)
& MVA $\leq$ 1.5 cm$^2$ (by planimetry, PHT, or continuity equation), or mean transmitral gradient $\geq$ 5--10 mmHg at resting heart rate, or PHT $\geq$ 150 ms \\
Mitral regurgitation (mod+)
& VC $\geq$ 3 mm, or EROA $\geq$ 0.20 cm$^2$ (primary MR; $\geq$ 0.20 cm$^2$ also used for severe secondary MR per some criteria), or RVol $\geq$ 30 mL, or RF $\geq$ 30\% \\
Tricuspid regurgitation (mod+)
& VC $\geq$ 3 mm, or EROA $\geq$ 0.20 cm$^2$, or RVol $\geq$ 30 mL, or PISA radius $>$ 6 mm (Nyquist $\sim$28 cm/s) \\
\bottomrule
\end{tabular}%
}
\par\vspace{3pt}
\noindent\parbox{\textwidth}{\footnotesize
\textbf{Note:} SHD, structural heart disease; LV, left ventricular; LVEF, left ventricular ejection fraction; LVIDd, left ventricular internal diameter in diastole; EDV, end-diastolic volume; BSA, body surface area; LA, left atrial; LAD, left atrial diameter; RV, right ventricular; RVD1, right ventricular basal diameter; RVD2, right ventricular mid-cavity diameter; RA, right atrial; VC, vena contracta; PHT, pressure half-time; $V_{\max}$, maximum velocity; EROA, effective regurgitant orifice area; AVA, aortic valve area; PISA, proximal isovelocity surface area; MVA, mitral valve area.}
\end{table*}

\subsection*{Data sources and cohort construction}

This is a retrospective, multicenter diagnostic accuracy study with an exploratory longitudinal sub-analysis. This study utilized longitudinal ECG and transthoracic echocardiography data from two tertiary medical centers in China: Peking University People's Hospital (PKUPH; Beijing, development center) and the Second Hospital of Tianjin Medical University (SHTMU; Tianjin, external validation center). The external site provided a geographically independent cohort located approximately 140~km from the development site. The study protocol was approved by the Institutional Review Boards of PKUPH (No.~2024PHB428-001) and SHTMU (No.~KY2025K386). The requirement for informed consent was waived in accordance with relevant ethical guidelines.

\begin{figure*}[h]  
    \centering  
    \includegraphics[width=1.0\textwidth]{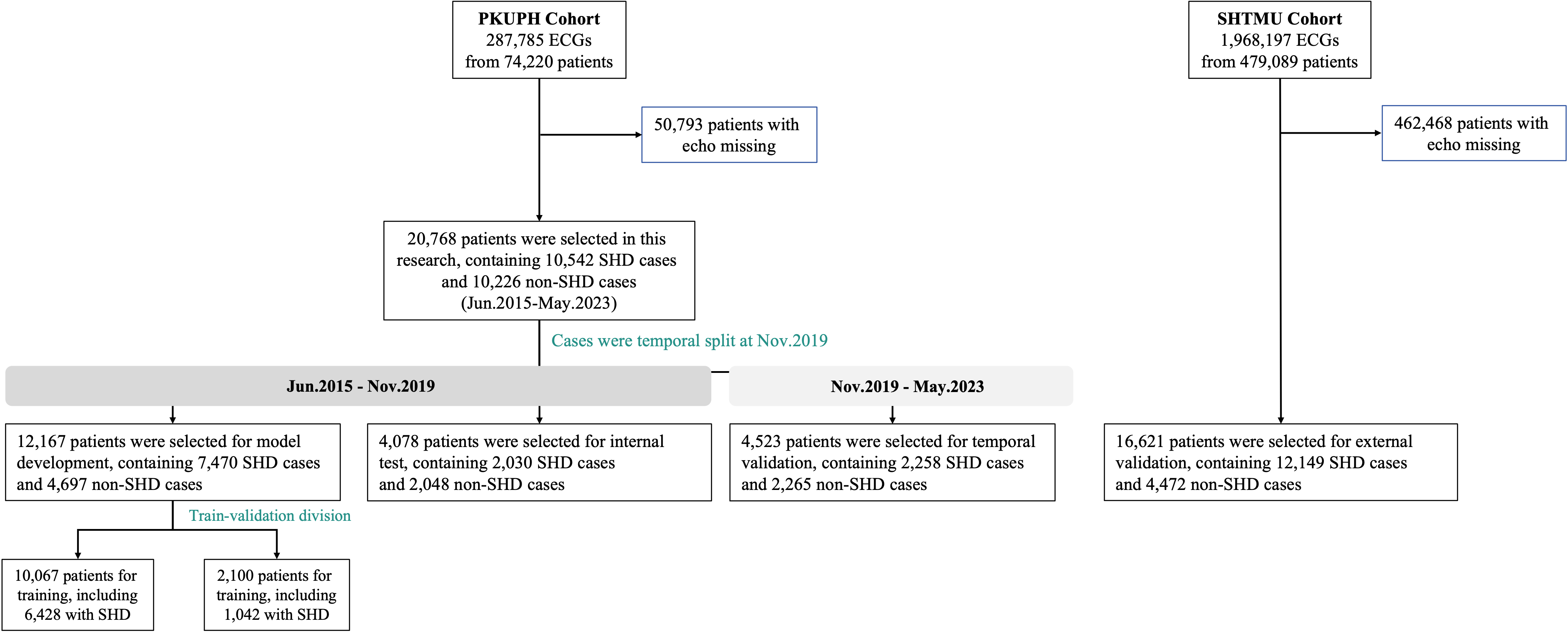}  
    \caption{Overview of the data selection procedure and cohort partitioning. ECG–echo pairs from PKUPH (Center~A) were temporally split at November 2019 into a development set (training and validation), an internal test set, and a temporal validation set. Cases from SHTMU (Center~B) served as an independent external validation cohort.}  
    \label{dataset_overview}  
\end{figure*}

Matched ECG--echocardiography pairs were identified at the patient level using a hierarchical temporal strategy. For each report, we prioritized same-day ECGs, expanding the search window up to $\pm$10 days to select the temporally closest recording. Pairs exceeding this 10-day window were excluded to minimize interval cardiac changes. Across both centers, this strategy yielded matched data for a total of 37,389 unique patients. To prevent data leakage, all partitions were conducted at the patient level, ensuring no individual's records overlapped between the training, validation, and testing sets (Fig.~\ref{dataset_overview}).

At PKUPH (June 2015--May 2023), 20,768 patients were identified from 74,220 screened individuals. To evaluate model robustness against distributional shifts, data were stratified temporally using a November 2019 cutoff. Patients from the earlier period were allocated to training ($n=10,067$), validation ($n=2,100$), and internal testing ($n=4,078$). Patients from the later period formed a distinct temporal validation set ($n=4,523$) to assess performance over time. At the external center, 16,621 patients were selected from an institutional repository of 479,089 individuals for geographically independent validation. Detailed cohort characteristics and allocation strategies are summarized in Fig.~\ref{dataset_overview}.

\subsubsection*{Data preprocessing}

All ECG recordings were processed through a standardised pipeline before being used for model development. Records containing corrupted waveforms, incomplete lead data, or inconsistent metadata were first discarded. ECG signals acquired at sampling rates below 500~Hz were upsampled to 500~Hz through linear interpolation to ensure a consistent temporal resolution across all recordings. A notch filter centred at 50/60~Hz was applied to each signal to attenuate power-line interference, after which a fourth-order Butterworth bandpass filter with a passband of 0.67--40~Hz was employed to retain physiologically meaningful frequency components while suppressing both high-frequency noise and low-frequency artefacts. Residual baseline wander was corrected by subtracting a median-filtered baseline estimate computed over a 0.4-second sliding window. The resulting signal was thus constrained to the 0.67--40~Hz frequency band with power-line interference and baseline drift minimised. For recordings exceeding 10 seconds in duration, a 10-second segment was extracted from the beginning of the signal; recordings shorter than 10 seconds were zero-padded to reach the target length. Prior to model input, each 10-second segment was normalised to zero mean and unit variance computed over that segment.

\subsection*{AnyECG-Echo architecture}

AnyECG-Echo is a deep learning framework designed for structural heart disease (SHD) detection from single-lead ECG. The framework operates in two stages: multimodal contrastive pre-training and supervised fine-tuning. During pre-training, the model aligns raw single-lead waveforms with physician-authored echocardiography reports in a shared latent space. In the fine-tuning stage, the pre-trained, frozen ECG encoder serves as a feature extractor for a linear classification head to detect 13 specific SHD subtypes.

\subsubsection*{Single-lead ECG encoder}

The ECG encoder maps a single-lead recording $\mathbf{X}_e \in \mathbb{R}^{1 \times L}$ ($L$ denotes temporal length) into a low-dimensional representation. We first employ a one-dimensional residual network (ResNet18) to extract hierarchical temporal features:
\begin{equation}
\mathbf{H} = f_{\text{ResNet}}(\mathbf{X}_e;\,\theta_{\text{res}}) \in \mathbb{R}^{C \times T}
\end{equation}
where $C$ and $T$ denote channel and downsampled temporal dimensions. A $1 \times 1$ convolution projects $\mathbf{H}$ into an embedding sequence $\mathbf{Z} \in \mathbb{R}^{d_{\text{proj}} \times T}$.

To capture global context, we process $\mathbf{Z}$ using a multi-head self-attention layer. A learnable class token [CLS] and positional embeddings are added, and the output corresponding to the [CLS] token, $\mathbf{e}_e \in \mathbb{R}^{d_{\text{out}}}$, serves as the global ECG embedding.

\subsubsection*{Contrastive pre-training objective}

The encoder is optimized using a weighted sum of unimodal and cross-modal contrastive losses: $\mathcal{L}_{\text{pre}} = \lambda_{\text{ECG}} \mathcal{L}_{\text{ECG}} + \lambda_{\text{EM}} \mathcal{L}_{\text{EM}}$.

Unimodal Loss ($\mathcal{L}_{\text{ECG}}$): To ensure representation invariance, we generate two augmented views, $\mathbf{z}_e^{(1)}$ and $\mathbf{z}_e^{(2)}$, via independent dropout and linear projections. We minimize an InfoNCE-style loss:
\begin{equation}
\mathcal{L}_{\text{ECG}} = - \frac{1}{N} \sum_{i=1}^{N} \log \frac{\exp\big(\mathrm{sim}(\mathbf{z}_{e,i}^{(1)}, \mathbf{z}_{e,i}^{(2)}) / \tau\big)}{\sum_{j=1}^{N} \exp\big(\mathrm{sim}(\mathbf{z}_{e,i}^{(1)}, \mathbf{z}_{e,j}^{(2)}) / \tau\big)}
\end{equation}

Cross-modal Loss ($\mathcal{L}_{\text{EM}}$): Single-lead ECG embeddings are aligned with semantic text embeddings $\tilde{\mathbf{e}}_t$ derived from frozen MedCPT query encoder~\cite{jin2023medcpt}. We employ a symmetric CLIP-style objective:
\begin{equation}
\mathcal{L}_{\text{EM}} = - \frac{1}{2N} \left( \sum_{i=1}^{N} \log \frac{\exp(s_{ii})}{\sum_{j=1}^{N} \exp(s_{ij})} + \sum_{j=1}^{N} \log \frac{\exp(s_{jj})}{\sum_{i=1}^{N} \exp(s_{ij})} \right)
\end{equation}
where $s_{ij}$ is the scaled cosine similarity between ECG $i$ and MedCPT-encoded report $j$.

\subsubsection*{Supervised fine-tuning}

For downstream detection, we append a linear classification head to the frozen single-lead ECG encoder. The average-pooled representation $\bar{\mathbf{z}}$ is mapped to predicted logits $\mathbf{o} \in \mathbb{R}^{M}$ ($M=13$):
\begin{equation}
\mathbf{o} = \mathbf{W}_{\text{cls}} \bar{\mathbf{z}} + \mathbf{b}_{\text{cls}}
\end{equation}
The classification head is trained using binary cross-entropy loss, while the encoder remains fixed to preserve the foundational physiological features learned during pre-training.


\section*{Data Availability}
The data that support the findings of this study are not publicly available due to restrictions imposed by institutional ethics committees and data governance policies of the participating hospitals. Access to the data may be considered upon reasonable request to the corresponding author, subject to approval by the relevant ethics committees.

\section*{Code Availability}
The code used for model development and evaluation in this study is publicly available at  \url{https://github.com/gody123gody/AnyECG-Echo}, or \url{https://github.com/PKUDigitalHealth/AnyECG-Echo}.

\section*{Acknowledgements}
This work is partially supported by the National Natural Science Foundation of China (62102008, 82470527), CCF-Tencent Rhino-Bird Open Research Fund (CCF-Tencent \\RAGR20250108), CCF-Zhipu Large Model Innovation Fund (CCF-Zhipu202414), PKU-OPPO Fund (BO202301, BO202503), Research Project of Peking University in the State Key Laboratory of Vascular Homeostasis and Remodeling (2025-SKLVHR-YCTS-02), Beijing Municipal Science and Technology Commission (Z251100000725008), Prevention and Control of Emerging and Major Infectious Diseases-National Science and Technology Major Project (2025ZD01906000, 2025ZD01906004).

\section*{Author Contributions}

Chenyang He and Qinghao Zhao contributed equally to this work. Kangyin Chen, Cheng Ding, and Shenda Hong are co-corresponding authors.

Chenyang He: Conceptualization, Methodology, Software, Formal analysis, Visualization, Writing -- original draft.
Qinghao Zhao: Data curation, Investigation, Validation, Writing -- review \& editing.
Shun Huang, Jun Li, and Gongzheng Tang: Software, Validation, Formal analysis.
Hao Zhang, Tong Liu, and Zhengkai Xue: Data curation, Resources.
Kangyin Chen: Resources, Data curation, Supervision -- review \& editing.
Cheng Ding: Methodology, Supervision -- review \& editing.
Shenda Hong: Conceptualization, Supervision, Project administration, Funding acquisition, Writing -- review \& editing.

\section*{Competing Interests}
The authors declare no competing interests.

\bibliography{reference}

\bigskip


\begin{appendices}
\onecolumn

\section*{Supplementary Table: Study population}

The study population comprised two major cohorts: a development cohort of 20,768 patients from Peking University People's Hospital (Beijing, China) and a geographically independent external validation cohort of 16,621 patients from the Second Hospital of Tianjin Medical University (Tianjin, China). Detailed baseline characteristics are provided in Table~\ref{tab:demographics}. 

\begin{table*}[h]
\centering
\caption{\textbf{Characteristics of multi-centre cohort.} LV/LA, left ventricular/atrial; RV/RA, right ventricular/atrial.}
\label{tab:demographics}
\small
\setlength{\tabcolsep}{10pt}
\renewcommand{\arraystretch}{1.05}
\resizebox{0.55\textwidth}{!}{%
\begin{tabular}{@{} l r r @{}}
\toprule
 & \textbf{PKUPH} & \textbf{SHTMU} \\
\midrule
\textbf{Patients ($n$)} & 20,768 & 16,621 \\
\textbf{ECGs ($n$)} & 27,158 & 18,588 \\
\textbf{Age, years (Mean $\pm$ SD)} & 61.6 $\pm$ 14.4 & 65.3 $\pm$ 13.7 \\
\textbf{Sex}, $n$ (\%) & & \\
\quad Male & 9,193 (45.3) & 9,654 (59.1) \\
\midrule
\textbf{SHD prevalence}, $n$ (\%) & & \\
\quad Reduced LV systolic function & 573 (2.8) & 1,478 (8.9) \\
\quad Pericardial effusion & 444 (2.1) & 67 (0.4) \\
\quad LV enlargement & 942 (4.5) & 26 (0.2) \\
\quad LA enlargement & 6,657 (32.1) & 10,087 (60.7) \\
\quad RV enlargement & 189 (0.9) & 41 (0.2) \\
\quad RA enlargement & 233 (1.1) & 99 (0.6) \\
\quad Global heart enlargement & 83 (0.4) & 289 (1.7) \\
\quad Ascending aortic dilation & 2,059 (9.9) & 1,808 (10.9) \\
\quad Aortic regurgitation & 138 (0.5) & 265 (1.6) \\
\quad Aortic stenosis & 73 (0.3) & 11 (0.1) \\
\quad Tricuspid regurgitation & 195 (0.7) & 535 (3.2) \\
\quad Mitral regurgitation & 294 (1.1) & 531 (3.2) \\
\quad Mitral stenosis & 69 (0.3) & 22 (0.1) \\
\bottomrule
\end{tabular}%
} 
\end{table*}

\section*{Supplementary Table: Performance benchmarking using 12-lead ECG configurations}

While the primary focus of this study is SHD screening via wearable-compatible single-lead ECGs, we evaluated the AnyECG-Echo framework using standard clinical 12-lead ECG inputs to establish a performance benchmark. As shown in Table~\ref{tab:main_results_12l}, the transition to a 12-lead configuration provides a modest but consistent improvement in discrimination across nearly all SHD categories. This gain is most notable in conditions where multi-vector electrophysiological information is critical, such as valvular regurgitation and right heart remodeling. Crucially, the framework maintains high external generalizability even with the increased complexity of 12-lead inputs. These findings confirm that AnyECG-Echo effectively scales to different clinical hardware environments while maintaining the same hierarchical diagnostic patterns observed in the single-lead analysis.

A comparison of Table~\ref{tab:main_results_1l} and Table~\ref{tab:main_results_12l} further reveals that single-lead performance retained a consistently high proportion of 12-lead AUROC across all labels.

\begin{table*}[t]
\centering
\caption{\textbf{Summary of AUROC performance} with 1-lead ECG (including Positive/Negative counts).}
\label{tab:main_results_1l}
\renewcommand{\arraystretch}{1.35}
\resizebox{\textwidth}{!}{%
\begin{tabular}{@{} l ccc ccc ccc @{}}
\toprule
 & \multicolumn{3}{c}{\textbf{Internal test}} 
 & \multicolumn{3}{c}{\textbf{Temporal validation}} 
 & \multicolumn{3}{c}{\textbf{External validation}} \\
\cmidrule(lr){2-4} \cmidrule(lr){5-7} \cmidrule(lr){8-10}
\textbf{Cardiac condition}
 & \textbf{N (P/N)} & \textbf{AUROC} & \textbf{95\% CI}
 & \textbf{N (P/N)} & \textbf{AUROC} & \textbf{95\% CI}
 & \textbf{N (P/N)} & \textbf{AUROC} & \textbf{95\% CI} \\
\midrule

Mitral stenosis
 & 8/4070 & 0.836 & [0.685, 0.971]
 & 11/4512 & 0.889 & [0.735, 0.977]
 & 22/16599 & 0.906 & [0.837, 0.951] \\

Global heart enlargement
 & 20/4058 & 0.931 & [0.880, 0.971]
 & 25/4498 & 0.929 & [0.883, 0.963]
 & 289/16332 & 0.877 & [0.860, 0.894] \\
 
Reduced LV systolic function
 & 110/3968 & 0.924 & [0.895, 0.950]
 & 139/4384 & 0.920 & [0.901, 0.936]
 & 1478/15143 & 0.866 & [0.858, 0.874] \\

 Tricuspid regurgitation
 & 28/4050 & 0.907 & [0.838, 0.962]
 & 32/4491 & 0.893 & [0.837, 0.936]
 & 535/16086 & 0.820 & [0.800, 0.837] \\

 Mitral regurgitation
 & 40/4038 & 0.808 & [0.739, 0.870]
 & 41/4482 & 0.842 & [0.786, 0.889]
 & 531/16090 & 0.806 & [0.788, 0.823] \\

RA enlargement
 & 36/4042 & 0.914 & [0.871, 0.952]
 & 66/4457 & 0.913 & [0.887, 0.936]
 & 99/16522 & 0.803 & [0.762, 0.843] \\

Pericardial effusion
 & 88/3990 & 0.728 & [0.676, 0.777]
 & 96/4427 & 0.691 & [0.646, 0.734]
 & 67/16554 & 0.832 & [0.790, 0.870] \\

RV enlargement
 & 42/4036 & 0.882 & [0.833, 0.930]
 & 35/4488 & 0.771 & [0.689, 0.846]
 & 41/16580 & 0.748 & [0.670, 0.819] \\
 
LV enlargement
 & 197/3881 & 0.868 & [0.843, 0.895]
 & 216/4307 & 0.838 & [0.812, 0.861]
 & 26/16595 & 0.698 & [0.605, 0.777] \\

 Aortic stenosis
 & 15/4063 & 0.835 & [0.734, 0.917]
 & 13/4510 & 0.756 & [0.616, 0.883]
 & 11/16610 & 0.714 & [0.576, 0.832] \\

Ascending aortic dilation
 & 342/3736 & 0.685 & [0.658, 0.713]
 & 507/4016 & 0.677 & [0.657, 0.700]
 & 1808/14813 & 0.635 & [0.622, 0.648] \\

Aortic regurgitation
 & 17/4061 & 0.928 & [0.841, 0.979]
 & 27/4496 & 0.807 & [0.721, 0.886]
 & 265/16356 & 0.639 & [0.609, 0.669] \\

LA enlargement
 & 1289/2789 & 0.665 & [0.648, 0.680]
 & 1406/3117 & 0.657 & [0.641, 0.673]
 & 10087/6534 & 0.610 & [0.602, 0.618] \\

\bottomrule
\end{tabular}%
}
\end{table*}

\begin{table*}[t]
\centering
\caption{\textbf{Summary of AUROC performance} with 12-lead ECG (including Positive/Negative counts).}
\label{tab:main_results_12l}
\renewcommand{\arraystretch}{1.35}
\resizebox{\textwidth}{!}{%
\begin{tabular}{@{} l ccc ccc ccc @{}}
\toprule
 & \multicolumn{3}{c}{\textbf{Internal test}} 
 & \multicolumn{3}{c}{\textbf{Temporal validation}} 
 & \multicolumn{3}{c}{\textbf{External validation}} \\
\cmidrule(lr){2-4} \cmidrule(lr){5-7} \cmidrule(lr){8-10}
\textbf{Cardiac condition}
 & \textbf{N (P/N)} & \textbf{AUROC} & \textbf{95\% CI}
 & \textbf{N (P/N)} & \textbf{AUROC} & \textbf{95\% CI}
 & \textbf{N (P/N)} & \textbf{AUROC} & \textbf{95\% CI} \\
\midrule

Mitral stenosis
 & 8/4070 & 0.972 & [0.947, 0.983]
 & 11/4512 & 0.944 & [0.930, 0.987]
 & 22/16599 & 0.916 & [0.882, 0.959] \\

Global heart enlargement
 & 20/4058 & 0.961 & [0.950, 0.980]
 & 25/4498 & 0.980 & [0.974, 0.987]
 & 289/16332 & 0.897 & [0.889, 0.904] \\
 
Reduced LV systolic function
 & 110/3968 & 0.937 & [0.929, 0.952]
 & 139/4384 & 0.930 & [0.921, 0.942]
 & 1478/15143 & 0.890 & [0.880, 0.895] \\

Tricuspid regurgitation
 & 28/4050 & 0.953 & [0.916, 0.973]
 & 32/4491 & 0.948 & [0.934, 0.964]
 & 535/16086 & 0.841 & [0.834, 0.851] \\

Mitral regurgitation
 & 40/4038 & 0.876 & [0.820, 0.920]
 & 41/4482 & 0.876 & [0.840, 0.908]
 & 531/16090 & 0.820 & [0.810, 0.835] \\

RA enlargement
 & 36/4042 & 0.916 & [0.911, 0.943]
 & 66/4457 & 0.922 & [0.902, 0.943]
 & 99/16522 & 0.803 & [0.752, 0.829] \\

Pericardial effusion
 & 88/3990 & 0.822 & [0.799, 0.857]
 & 96/4427 & 0.784 & [0.762, 0.811]
 & 67/16554 & 0.766 & [0.687, 0.778] \\

RV enlargement
 & 42/4036 & 0.932 & [0.907, 0.955]
 & 35/4488 & 0.754 & [0.704, 0.798]
 & 41/16580 & 0.735 & [0.655, 0.779] \\

LV enlargement
 & 197/3881 & 0.892 & [0.878, 0.907]
 & 216/4307 & 0.861 & [0.849, 0.887]
 & 26/16595 & 0.730 & [0.686, 0.794] \\

Aortic stenosis
 & 15/4063 & 0.867 & [0.771, 0.943]
 & 13/4510 & 0.707 & [0.602, 0.793]
 & 11/16610 & 0.700 & [0.650, 0.828] \\
 
Ascending aortic dilation
 & 342/3736 & 0.725 & [0.710, 0.744]
 & 507/4016 & 0.713 & [0.697, 0.726]
 & 1808/14813 & 0.674 & [0.668, 0.685] \\

Aortic regurgitation
 & 17/4061 & 0.956 & [0.938, 0.968]
 & 27/4496 & 0.782 & [0.707, 0.833]
 & 265/16356 & 0.673 & [0.655, 0.686] \\
 
LA enlargement
 & 1289/2789 & 0.725 & [0.718, 0.737]
 & 1406/3117 & 0.719 & [0.711, 0.724]
 & 10087/6534 & 0.642 & [0.637, 0.650] \\

\bottomrule
\end{tabular}%
}
\end{table*}

\section*{Supplementary Table: Subgroup Analysis of Low-Prevalence SHD Subtypes}
Supplementary Table~\ref{tab:renming_subgroup_template_remaining} provides a comprehensive demographic breakdown for the remaining SHD subtypes in the external validation cohort. Due to the low prevalence of certain conditions within specific age and sex strata, several categories contained fewer than five positive cases and are consequently reported as N/A to ensure statistical reliability.

Despite the limited sample sizes in these sub-populations, AnyECG-Echo maintained robust discriminative performance across the calculable categories. Notably, for RV enlargement and Mitral stenosis, the model achieved AUROCs consistently exceeding 0.80 across most age groups, further demonstrating the framework's ability to capture specific structural signatures even for less frequent pathologies. The performance across sex and age strata, where data were sufficient for calculation, remained generally consistent with the primary findings, suggesting that the model's diagnostic utility extends to a broad spectrum of valvular and chamber-specific abnormalities without manifesting significant demographic bias.

\begin{table*}[t]
\centering
\caption{\textbf{Subgroup analysis of the remaining labels in the PKUPH cohort, stratified by age group and sex}. Values are AUROC (95\% confidence interval). N/A indicates fewer than five positive cases in that specific stratum.}
\label{tab:renming_subgroup_template_remaining}
\setlength{\tabcolsep}{4pt}
\renewcommand{\arraystretch}{1.35}
\resizebox{\textwidth}{!}{%
\begin{tabular}{@{} l r cccccccc @{}} 
\toprule

\textbf{Subgroup}
    & \makecell[r]{\textbf{\textit{n}}\\(pos/neg)}
    & \makecell{\textbf{LA}\\\textbf{enlargement}}
    & \makecell{\textbf{RA}\\\textbf{enlargement}}
    & \makecell{\textbf{RV}\\\textbf{enlargement}}
    & \makecell{\textbf{Aortic}\\\textbf{regurg.}}
    & \makecell{\textbf{Aortic}\\\textbf{stenosis}}
    & \makecell{\textbf{Tricuspid}\\\textbf{regurg.}}
    & \makecell{\textbf{Mitral}\\\textbf{regurg.}}
    & \makecell{\textbf{Mitral}\\\textbf{stenosis}} \\

\midrule

\rowcolor[gray]{0.95}
\multicolumn{10}{@{}l}{\textbf{Age group (years)}} \\ 

$<$50    & 142/280
  & 0.617 [0.541, 0.688]
  & N/A
  & 0.856 [0.663, 0.981]
  & N/A
  & N/A
  & N/A
  & 0.787 [0.548, 0.963]
  & N/A \\

50--64    & 347/410
  & 0.644 [0.600, 0.689]
  & N/A
  & 0.917 [0.792, 0.998]
  & 0.814 [0.515, 0.984]
  & N/A
  & 0.840 [0.672, 0.990]
  & 0.708 [0.553, 0.856]
  & 0.845 [0.640, 0.993] \\

65--74    & 342/248
  & 0.634 [0.588, 0.680]
  & N/A
  & 0.953 [0.902, 0.992]
  & N/A
  & N/A
  & 0.973 [0.920, 0.999]
  & 0.880 [0.798, 0.949]
  & N/A \\

$\geq$75    & 267/136
  & 0.640 [0.587, 0.694]
  & 0.837 [0.681, 0.949]
  & N/A
  & N/A
  & N/A
  & N/A
  & N/A
  & N/A \\

\midrule

\rowcolor[gray]{0.95}
\multicolumn{10}{@{}l}{\textbf{Sex}} \\ 

Male     & 858/571
  & 0.641 [0.611, 0.670]
  & 0.868 [0.760, 0.948]
  & 0.906 [0.820, 0.976]
  & 0.844 [0.545, 0.977]
  & 0.903 [0.754, 0.978]
  & N/A
  & 0.838 [0.732, 0.923]
  & N/A \\

Female     & 753/999
  & 0.677 [0.649, 0.705]
  & 0.928 [0.867, 0.974]
  & 0.883 [0.802, 0.948]
  & N/A
  & 0.846 [0.734, 0.942]
  & 0.901 [0.812, 0.968]
  & 0.731 [0.598, 0.856]
  & 0.830 [0.673, 0.983] \\

\bottomrule
\multicolumn{10}{@{}p{1.88\linewidth}}{\footnotesize 
  \textit{n} = positive / negative ECG recordings per subgroup (SHD overall).
  AUROC values: point estimate [95\% confidence interval], calculated via 2,000 bootstrap iterations.
  RV, Right ventricular; LA, Left atrial; RA, Right atrial; Regurg., Regurgitation.}
\end{tabular}%
}
\end{table*}

\section*{Supplementary Table: Community prevalence estimates used for NNS calculation}

Community prevalence estimates were derived from published population-based echocardiographic screening studies. Where contemporary data were unavailable, we adopted conservative estimates from earlier landmark studies. Given the well-documented increase in SHD prevalence driven by population ageing~\cite{mensah2023global}, these figures likely underestimate the true contemporary prevalence, rendering our NNS estimates conservative.

\begin{table*}[h]
\centering
\caption{\textbf{Community prevalence estimates for SHD subtypes used in number-needed-to-screen (NNS) calculations}. Estimates reflect the general adult population unless otherwise noted.}
\label{tab:community_prevalence}
\renewcommand{\arraystretch}{1.35}
\resizebox{\textwidth}{!}{%
\begin{tabular}{@{} l c l l @{}}
\toprule
\textbf{Cardiac condition} & \textbf{Assumed prevalence} & \textbf{Reported range} & \textbf{Source} \\
\midrule
Reduced LV systolic function & 3.0\% & 2--6\% & Mosterd et al., \textit{Eur Heart J} 1999~\cite{mosterd1999prevalence}; Raymond et al., \textit{Heart} 2003~\cite{raymond2003prevalence} \\
Pericardial effusion & 1.0\% & 1--3\% & Morbach et al., \textit{JAHA} 2024~\cite{sahiti2024prognostic} \\
Global heart enlargement & 1.0\% & 0.4--1\%$^\dagger$ & Estimated from clinical series \\
Left atrial enlargement & 5.0\% & 2--10\% & Ou et al., \textit{BMC Cardiovasc Disord} 2016~\cite{ou2016prevalence} \\
LV enlargement & 2.0\% & 1--3\%$^\dagger$ & Vasan et al., \textit{Circulation} 1997~\cite{vasan1997left} \\
Right atrial enlargement & 1.0\% & $\sim$1\%$^\dagger$ & Estimated from clinical series \\
RV enlargement & 1.0\% & $\sim$1\%$^\dagger$ & Estimated from clinical series \\
Ascending aortic dilation & 3.0\% & 1.5--5\% & Nkomo et al., \textit{The lancet} 2006~\cite{nkomo2006burden}; d'Arcy et al., \textit{Eur Heart J} 2016~\cite{d2016large} \\
Aortic regurgitation (mod--sev) & 1.5\% & 1--2\% & Nkomo et al., \textit{The lancet} 2006~\cite{nkomo2006burden}; d'Arcy et al., \textit{Eur Heart J} 2016~\cite{d2016large} \\
Aortic stenosis (mod--sev) & 1.5\% & 1--2\% & Nkomo et al., \textit{The lancet} 2006~\cite{nkomo2006burden}; d'Arcy et al., \textit{Eur Heart J} 2016~\cite{d2016large} \\
Tricuspid regurgitation (mod--sev) & 1.0\% & 0.5--1\% & Nkomo et al., \textit{The lancet} 2006~\cite{nkomo2006burden}; d'Arcy et al., \textit{Eur Heart J} 2016~\cite{d2016large} \\
Mitral regurgitation (mod--sev) & 2.0\% & 1--2\% & Nkomo et al., \textit{The lancet} 2006~\cite{nkomo2006burden}; d'Arcy et al., \textit{Eur Heart J} 2016~\cite{d2016large} \\
Mitral stenosis (mod--sev) & 0.5\% & 0.1--1\%$^\ddagger$ & Nkomo et al., \textit{The lancet} 2006~\cite{nkomo2006burden}; d'Arcy et al., \textit{Eur Heart J} 2016~\cite{d2016large} \\

\bottomrule
\end{tabular}%
}
\vspace{4pt}
\raggedright
\footnotesize{$^\dagger$No large-scale population-based echocardiographic data available; estimate derived from disease-specific prevalence studies and clinical series.}\\
\footnotesize{$^\ddagger$Prevalence varies substantially by geography; rheumatic mitral stenosis is rare in high-income countries ($<$0.1\%) but more common in low- and middle-income settings (1--2\%).}
\end{table*}

\section*{Supplementary Figure: Electrophysiological interpretation results}
\begin{figure*}[t]
    \centering
    \includegraphics[width=1.0\textwidth]{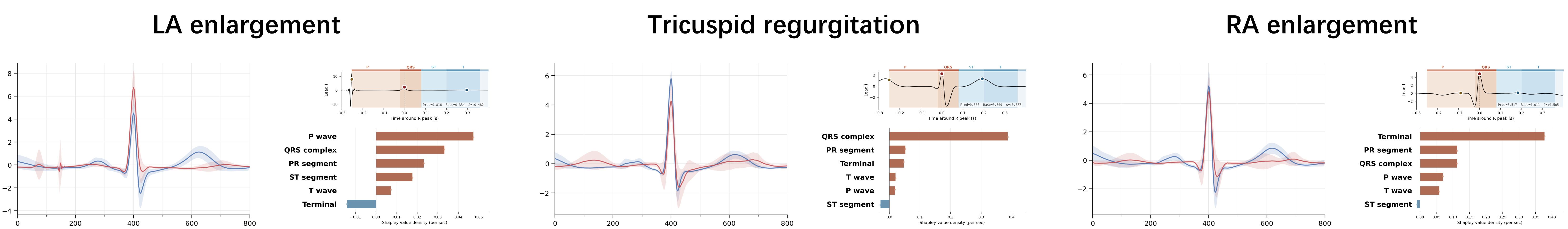}
\caption{\textbf{Supplementary Electrophysiological interpretation of learned ECG features across SHD phenotypes.}}
    \label{fig:morphology_2}
\end{figure*}

A distinct set of results characterizing the model’s sensitivity to subtle atrial signatures is summarized in Fig.~\ref{fig:morphology_2}. While AnyECG-Echo identifies P-mitrale patterns in mitral stenosis through combined P-wave and QRS complex attribution, its capture of isolated P-wave morphology remains incomplete. For conditions such as tricuspid regurgitation or right atrial enlargement, the P-wave broadening and amplitude shifts in Lead I are often too subtle for robust single-beat encoding. These findings suggest a technical nuance: the model’s discrimination of atrial-related labels likely relies more heavily on co-occurring ventricular or hemodynamic features than on direct P-wave signatures. This limitation reflects the inherent constraints of single-lead, single-beat analysis in resolving subtle atrial structural changes.

\end{appendices}

\end{document}